# ALICE Diffractive Detector Control System for RUN-II in the ALICE Experiment

**J C Cabanillas[1], M I Martínez[2], I León[1]**

[1] Universidad Autónoma de Sinaloa (UAS), Josefa Ortíz de Dominguez S/N Culiacan Rosales, Sinaloa, México, CP: 80013
[2] Benemérita Universidad Autónoma de Puebla (BUAP), Cd. Universitaria, 72570 Heróica Puebla de Zaragoza, Puebla, México

Email: juan.carlos.cabanillas.noris@cern.ch, mario.martinez.hernandez@cern.ch, ildefonso.leon.monzon@cern.ch

**Abstract**. The ALICE Diffractive (AD0) detector has been installed and commissioned for the second phase of operation (RUN-II). With this new detector it is possible to achieve better measurements by expanding the range of pseudo-rapidity in which the production of particles can be detected. Specifically the selection of diffractive events in the ALICE experiment which was limited by the range over which rapidity gaps occur. Any new detector should be able to take data synchronously with all other detectors and to be operated through the ALICE central systems. One of the key elements developed for the AD0 detector is the Detector Control System (DCS). The DCS is designed to operate safely and correctly this detector. Furthermore, the DCS must also provide optimum operating conditions for the acquisition and storage of physics data and ensure these are of the highest quality.

The operation of AD0 implies the configuration of about 200 parameters, as electronics settings and power supply levels and the generation of safety alerts. It also includes the automation of procedures to get the AD0 detector ready for taking data in the appropriate conditions for the different run types in ALICE. The performance of AD0 detector depends on a certain number of parameters such as the nominal voltages for each photomultiplier tube (PMT), the threshold levels to accept or reject the incoming pulses, the definition of triggers, etc. All these parameters affect the efficiency of AD0 and they have to be monitored and controlled by the AD0 DCS.

## 1. Introduction

ALICE (A Large Ion Collider Experiment) is an experiment dedicated to the study of strongly interacting matter, particularly to the study of heavy ion collisions at very high energies at the LHC in the European Organization for Nuclear Research (CERN) [1]. The new AD0 detector improves the detection capacity for diffractive events in proton-proton (p-p) and lead ion (Pb-Pb) collisions by adding four planes of particle counters at very small angles with respect to the beam direction. The detection of diffractive events in ALICE during the first phase of LHC operation (RUN-I) was limited by the range over which rapidity gaps were identified. AD0 will expand the range in which the production of particles (or lack thereof) can be detected [2]. Another benefit of the AD0 diffractive physics trigger is the larger amount of data recorded due to the fact that about 25% of the results of collision events are of the diffractive type [3]. To integrate AD0 to the ALICE data taking, it was necessary to design a new control architecture specific to this detector.

## 2. The AD0 Detector

The AD0 detector consists of two sub-detectors called ADA and ADC. Each of them comprises two detector layers, each one formed by four scintillator modules arranged around the LHC beam pipe. The ADA and ADC designations refer to the positions where they are installed at both ends of the ALICE experimental site with respect to the interaction point (IP). The positions of ADA and ADC in the ALICE reference frame and the nomenclature for layers and modules are shown in Fig. 1.

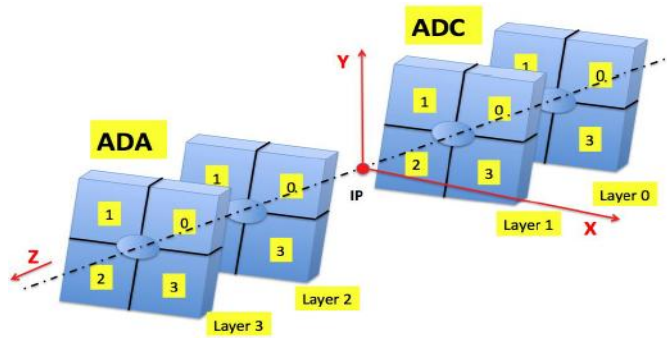

**Figure 1.** ADA and ADC sub-detectors nomenclature in the ALICE cavern

## 3. Detector Control System (DCS)

The DCS is responsible for controlling, monitoring and configuring the AD0 detector equipment among which there are commercial hardware devices like power supplies, voltage cards, crates, etc. as well as specific custom equipment like the Front-End Electronics (FEE). Also all relevant parameters for the offline analysis of physical data are periodically archived by the AD0 DCS in the ALICE Online Conditions Database (OCDB) [1]. These tasks are accomplished by sending commands and reading the status from the equipment. The control systems in ALICE are developed using a Supervisory Control and Data Acquisition (SCADA) platform called WinCC Open Architecture (OA) ® (formerly PVSS) from ETM Company. The AD0 DCS is integrated to the global control system of the ALICE experiment [4].

*3.1 AD0 DCS Software Architecture*
The software architecture is a tree-like hierarchy that models the structure of the hardware sub-systems and devices. This tree structure is composed of nodes, each one having a single parent, except the top node. The performance and functionality of each node in the tree hierarchy are implemented as a Finite State Machine (FSM) [5-6]. Fig. 2 shows the simplified software architecture in the AD0 DCS where the main subsystems are shown.

*3.2 AD0 DCS Hardware Architecture*
AD0 has an architecture compatible with other hardware architectures in the ALICE experiment which are subdivided into three layers; a) supervision, b) process control and c) field layer [1]. The AD0 detector hardware architecture is described in Fig. 3.

*3.3 Implementation of AD0 DCS*
The Control Unit (CU) and Device Unit (DU) nodes in the control hierarchy of AD0 are implemented as finite state machines. The FSM tool built into the DCS software is based on the SMI++ (State Machine Interface) language. The FSM Framework component allows the implementation of the DCS as a collection of logical objects (CU's) whose state depends on the state of the objects below in the control hierarchy. The state of the DU's depends on the real status of the hardware devices. On the opposite way commands are sent down the hierarchy up to the DU's who drive the devices to actually execute the commands [1].

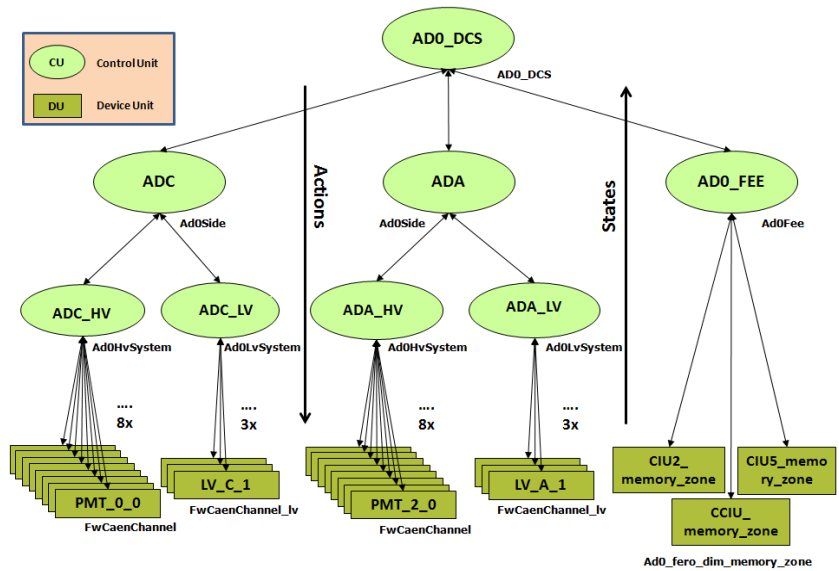

**Figure 2.** AD0 DCS Software Architecture.

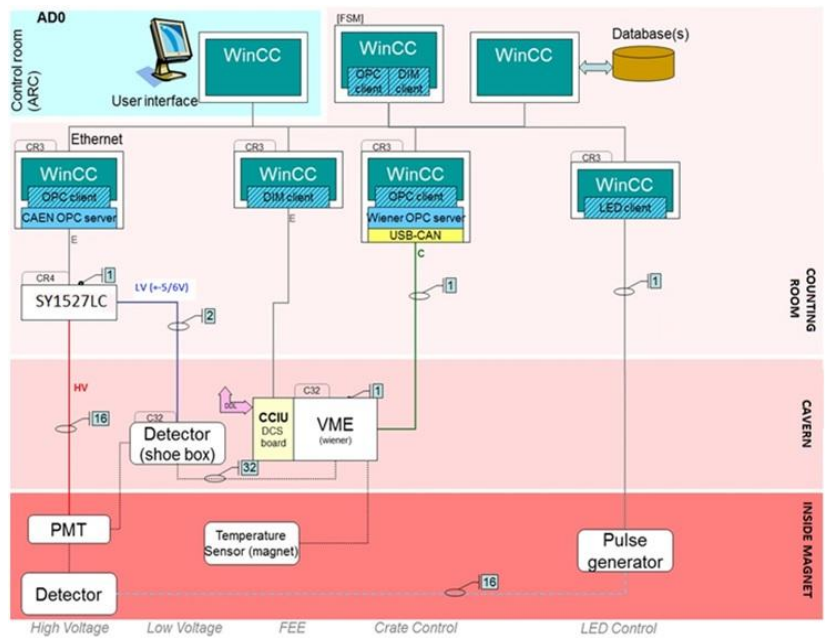

**Figure 3.** AD0 DCS Hardware Architecture.

*3.4 FSM Node on the Top Level of the Hierarchy*
At the highest level of the hierarchy the top node (AD0_DCS in Figure 2) is the main control unit. The FSM state diagram for this is shown in Fig. 4. This implementation of AD0 top node is based on the guidelines provided by the ALICE Control Coordination (ACC).

*3.5 DCS User Interface (UI)*
Custom user interfaces (panels) are associated with any CU or DU in the hierarchy. Commands can be sent, and states can be shown graphically from these interfaces. The user can also navigate through the hierarchy and display the operation panel for each node. The user interface of the AD0 DCS top node is shown in Fig. 5.

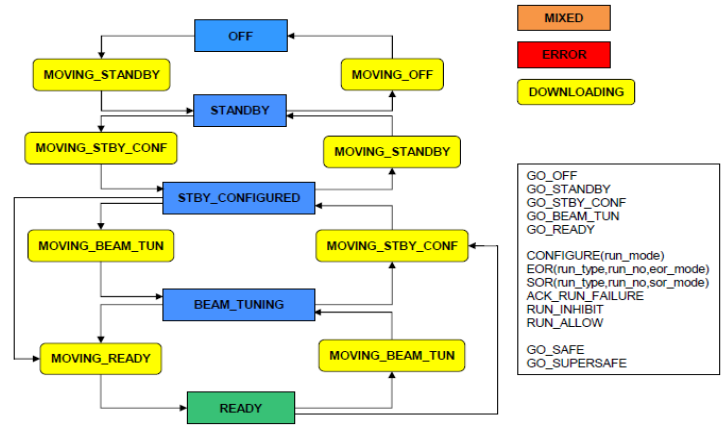

**Figure 4.** FSM state diagram of the AD0 top node.

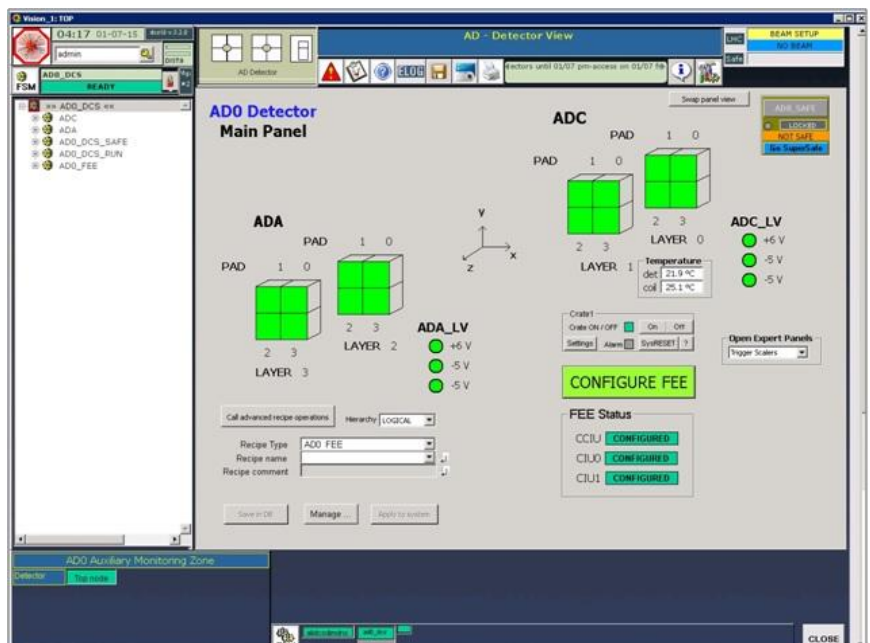

**Figure 5.** AD0 DCS main user interface.

## 4. Conclusions

The control system for AD0 detector was successfully developed and integrated in ALICE. AD0 detector is operational for data taking since March 2015 for the second running period (RUN II) of the LHC. AD0 DCS has standard and easy to operate interfaces for the detector expert. Also this control system is continuously maintained and updated following ALICE Control Coordination (ACC) requirements. Finally, AD0 control system has a stable operation within the ALICE central DCS.